\documentstyle[12pt,epsfig]{article}        
\newlength{\dinwidth}                       
\newlength{\dinmargin}                      
\def\lsim{\mathrel{\rlap{\lower4pt\hbox{\hskip1pt$\sim$}}
    \raise1pt\hbox{$<$}}}                
\def\gsim{\mathrel{\rlap{\lower4pt\hbox{\hskip1pt$\sim$}}
    \raise1pt\hbox{$>$}}}                
\newcommand{\bq}{\begin{equation}}
\newcommand{\eq}{\end{equation}}

\newcommand{\vph}{\vphantom{a long arbitary text: bla-bla-bla-bla-bla-bla}}
\begin{document}
\begin{titlepage}

\large
\normalsize
\begin{flushleft}
DESY 96--198\\
{\tt hep-ph/9609399}\\
September 1996
\end{flushleft}
\vspace*{4cm}
\begin{center}
\LARGE
{\bf Recent Developments in Radiative Corrections at HERA}\\

\vspace{3cm}
\large
Dima Bardin$^{a,b}$,
Johannes Bl\"umlein$^a$,
Penka Christova$^c$, \\
and Lida Kalinovskaya$^b$
\\
\vspace{1cm}
\large {\it
 $^a$DESY--Zeuthen \\

\vspace{0.1cm}
Platanenallee 6, D--15735 Zeuthen, Germany }\\

\vspace{0.1cm}
\large {\it
 $^b$Bogoliubov Laboratory for Theoretical Physics \\

\vspace{0.1cm}
JINR, ul. Joliot--Corie 6, RU--141980 Dubna, Russia}\\

\vspace{0.1cm}
\large {\it
 $^c$Konstantin Preslavsky University of Shoumen\\

\vspace{0.1cm}
9700 Shoumen, Bulgaria }\\
\vspace{1cm}
\normalsize
{\bf Abstract} \\
\end{center}
\noindent
\vfill
\noindent
\normalsize
We describe several numerical results for
radiative corrections for deep
inelastic $ep$ scattering at HERA which
are calculated using the {\tt HECTOR} package.
We present radiative corrections
for ten different choices of kinematical
variables for unpolarized neutral and charged current
deep inelastic scattering.
Radiative corrections for neutral current
scattering off polarized protons are calculated
in leptonic variables and
compared to those obtained by the
{\tt POLRAD} code for the kinematic regime
of the HERMES experiment.
\begin{center}
{
\small
\sf
Contribution to the Proceedings of the 1996 HERA Physics
Workshop}
\end{center}

\end{titlepage}

\vspace*{1cm}
\begin{center}  \begin{Large} \begin{bf}
     Recent Developments in Radiative Corrections at HERA \\
  \end{bf}  \end{Large}
  \vspace*{5mm}
  \begin{large}

Dima Bardin$^{a,b}$,~Johannes Bl\"umlein$^a$,~Penka Christova$^c$ \\
and~Lida Kalinovskaya$^b$ 
  \end{large}

$^a$ DESY--Zeuthen,
     Platanenallee~6,~D-15735~Zeuthen, Germany\\
$^b$ Bogoliubov Laboratory for Theoretical Physics, \\
JINR, ul. Joliot-Curie 6, RU--141980 Dubna, Russia \\
$^c$ Konstantin Preslavsky University of Shoumen,
9700 Shoumen, Bulgaria    \\
\end{center}
\begin{quotation}
\noindent
{\bf Abstract:}
We describe several numerical results for
radiative corrections for deep
inelastic $ep$ scattering at HERA which
are calculated using the {\tt HECTOR} package.
We present radiative corrections
for ten different choices of kinematical
variables for unpolarized neutral and charged current
deep inelastic scattering.
Radiative corrections for neutral current
scattering off polarized protons are calculated
in leptonic variables and
compared to those obtained by the
{\tt POLRAD} code for the kinematic regime
of the HERMES experiment.
\end{quotation}

\section{Introduction}
\label{sect1}

The precise knowledge of QED radiative corrections is indispensable
in the determination of nucleon structure functions. The forthcoming
high statistics measurements of $F_2(x,Q^2)$, $F_L(x,Q^2)$ and
$F^{c\bar{c}}(x,Q^2)$ at H1 and ZEUS require knowing the radiative corrections
at the $\%$ level. In some of the measurements particularly, the range
of high $y$ is essential. Here the radiative corrections turn out to be
large for some choices of the kinematical variables and higher order
corrections can be necessary.

In the present note we summarize the status reached in the calculation
of the QED radiative corrections. 
In section 2, we present a short
description of main features of the recently released code {\tt HECTOR}~\cite{he}.

For the first time also polarized nucleon
structure functions can be measured at HERA
with the HERMES experiment
Also here the radiative corrections
are large. Recently a dedicated new calculation~\cite{ph} was performed
including both $\gamma$ and $Z$-boson exchange 
and accounting for all twist-2 contributions
to the structure functions contributing
to scattering cross sections both for the case of
longitudinally and transverse polarized nucleons.

In section 3, we present a discussion of numerical results
summarized in a collection of figures.
A particular emphasis is given on the high $y$
range by presenting and discussing the results for ten different
choices of the kinematical variables.
A first comparison between the results of {\tt HECTOR}
and earlier results of
{\tt POLRAD}~\cite{ks}-\cite{pr} is presented.

\section{{\tt HECTOR 1.00} and its recent upgrade}

The code {\tt HECTOR} was created at DESY--Zeuthen in 1995.
Version {\tt 1.00}, November 1995~\cite{he},
accumulates and comprises results collected over the course of 20 years
(1975-1994) by the Dubna-Zeuthen Radiative Correction Group (DZRCG)~\cite{dz},
based on a {\it semi-analytic, model-independent} (MI) approach
and results by J.~Bl\"umlein (1990-1994),
based on an inclusive {\it leading logarithmic approach} (LLA).

\vspace{0.5cm}

\noindent
The branches of~{\tt HECTOR} include earlier codes for treatment
of selected parts of radiative corrections:

\begin{itemize}

\item
{\tt HELIOS} -- an inclusive LLA treatment of {\it leptonic} QED
radiative corrections including
second order initial state radiation,
${\cal{O}}((\alpha L)^2)$,
and {\it soft-photon exponentiation} to all orders
for a variety of measurements:
leptonic, mixed, Jaquet-Blondel, double angle variables,
the $\Sigma$ method and others~\cite{ll};

\item
{\tt TERAD} -- a complete ${\cal{O}}(\alpha)$
MI treatment of {\it leptonic} QED
radiative corrections for several types of measurements,
for a detailed description see~\cite{mi};

\item
{\tt DISEP} -- a complete ${\cal{O}}(\alpha)$
quark-parton model
treatment of QED radiative corrections
and {\it one-loop electroweak}
radiative corrections~\cite{qp} for leptonic and mixed
variables;

\item
{\tt TERADLOW} -- a MI treatment of {\it leptonic} QED
radiative corrections
in the photoproduction region for leptonic variables~\cite{mi}.

\end{itemize}

{\tt HECTOR} makes use of
extensive access to existing libraries of the structure functions
and parton densities,
both via the PDF-library~\cite{pd} and directly, and to recent
{\it low} $Q^2$-libraries.

The QCD corrections are implemented in
the framework of different factorization
schemes, as the $\overline{MS}$ and $DIS$ schemes, in order
to ensure a proper use of available parton densities.
The $LO$ option is also available.

Currently one may
access ten different choices of kinematical variables for
neutral and charged current deep inelastic scattering,
see figures 2a-j.

Simple kinematical cuts are possible
within the complete ${\cal{O}}(\alpha)$ MI approach.

\vspace{0.5cm}

\noindent
The upgrade of~{\tt HECTOR}, version {\tt 1.11}, will
contain the following additions:

\begin{itemize}

\item
The option to calculate radiative corrections for
neutral current deep inelastic polarized lepton -- polarized nucleon
scattering has been
incorporated~\cite{ph}. It includes both $\gamma$ and $Z$-boson exchange.
The Born cross section contains all twist-2 contributions to the polarized
structure functions -- for the cases of longitudinal and transverse proton
polarization.

\item
The radiative corrections
for  a {\it tagged photon measurement}~~based on a mixture of
complete MI, deterministic and LLA approaches
are being incorporated~\cite{tp}.

\end{itemize}

\noindent
Version~{\tt 1.11}
will be released by the end of 1996. 

\section{Numerical Results}
\label{sect3}

\subsection{QED radiative corrections at high $y$}

The radiative corrections (RC)
at high $y$ are presented in two sets of figures, 1 and 2,
at HERA collider energies. Here for the structure functions, we used
the CTEQ3M LO parametrization~\cite{3m}.

\vspace{0.25cm}

In figures 1a-d, we show the comparison between complete
${\cal{O}}(\alpha)$ MI calculations and those in LLA, for 4 types of
measurements for which the complete results are available.

In {\it leptonic} variables at small $x$ and high $y$,
where the correction is big, the difference between complete and LLA
calculations reaches tens of percent.

In {\it mixed} variables we registered an almost constant, $x,y$-independent
shift between the two calculations, which is quite small, $\leq 0.5\%$.

An interesting phenomenon is observed in {\it Jaquet-Blondel} and {\it hadronic}
variables. There the difference between the two calculations grows with
growing $y$, reaching several percent for $y\approx 1$, i.e. in the soft photon
corner of hadronic $y$. This could be a
reflection of the fact that in these variables the final state radiation
leading log correction is absent and non-logarithmic terms can be important.

So, one can conclude, that although
the LLA approximates the gross features of radiative corrections
in all 4 variables,
its precision is not sufficient if one aims at an accuracy
of measurement of the order of $1\%$.

\vspace{0.25cm}

In figures 2a-j, we show the comparison between
lowest order and higher order LLA calculations
of radiative corrections
for ten measurements: eight -- for neutral current (NC)
and two -- for charged current (CC).
Although we have presented figures for all ten choices of measurements
available in {\tt HECTOR}, we will discuss only several of the most popular
kinematic variables.

In {\it leptonic} variables at small $x$ and high $y$,
the higher order corrections reach tens of percent. Since LLA qualitatively
decribes the lowest order corrections, one may trust the reliability of
higher order corrections estimation.

In {\it Jaquet-Blondel}, {\it mixed}, and {\it hadronic}
variables, the higher order corrections exhibit very similar properties.
They grow with increasing $x$ and $y$,
reaching $1-2\%$ at high $x$ and high $y$,
i.e. in the soft photon corner.

The constant positive shift, growing with increasing $x$, 
is distinctly seen in {\it double angle} and ${\Sigma}$ variables.
It may reach $1\%$ for $x=0.1$ and goes down rapidly with decreasing $x$.

In the ${e\Sigma}$ method, the higher order corrections are
surprisingly large, but this
method is not so popular.

From figures presented, we may conclude that higher order corrections
are in general
rather important for the precision measurement of deep inelastic scattering
at HERA.

\vspace{0.25cm}

Both sets of figures prove that a realistic
radiative corrections
procedure must take into account
both the complete lowest order calculations and
higher order corrections, at least within LLA.
This is exactly the strategy that
the {\tt HECTOR} code follows.

\subsection{Comparison of~~{\tt HECTOR} and~~{\tt POLRAD15}
for Polarized Deep Inelastic Scattering}

\noindent
In this section we compare
the results of the codes {\tt HECTOR} and {\tt POLRAD15}.
We refer to the kinematic range of the HERMES experiment and consider only
leptonic corrections in leptonic variables for scattering
off polarized protons.
Both the cases of
longitudinal and transverse polarizations were studied.
We used the parametrizations of
Sch\"afer'88~\cite{88}
and GRSV'96~\cite{96} to describe the polarized structure
functions.

To simplify the first comparison between the two codes, we neglect
$Z$-boson exchange, account only for
one structure function
($g^{\gamma\gamma}_1$) for the case
of longitudinal proton polarization, and
at most for two ($g^{\gamma\gamma}_{1,2}$)
for the case of transverse polarization.
Furthermore, we neglect the vacuum polarization correction
(the running $\alpha$), higher order radiative corrections,
hadronic corrections and electroweak corrections.

The comparison for {\it unpolarized, longitudinal} and {\it transverse} 
parts of deep inelastic scattering (DIS)
radiative corrections
normalized by corresponding Born cross sections
is presented in figures 3a-d.
They are denoted in figures
as {\small UNPOL,~LONG,~TRAN,} correspondingly.
Further details may be found in~\cite{ph}.

\vspace{5mm}

The results of the comparison may be summarized as follows:

\begin{itemize}
\item Very good agreement was found for all values of
      $x$ and $y$ in the {\it unpolarized}-case.
\item We registered some disagreement
      for small $x$ and high $y$, i.e. in the Compton region,
      for {\it longitudinal}- and {\it transverse}-cases.
\item We observed an amazing
      agreement between complete and LLA calculations in the considered
      set-up. This suggests the use of a fast, LLA code for HERMES
      measurements in leptonic variables.
\end{itemize}

\noindent
Further work on the comparison, aimed at the resolution of the above-mentioned
      disagreement, is in progress.

\subsection*{Acknowledgments}
The authors are very much obliged
to I.~Akushevich for pleasant and fruitful collaboration
on the comparison with {\tt POLRAD}.
We are thankful to A.~Arbuzov and T.~Riemann
for an enjoyable common work within the {\tt HECTOR} project.
We are indepted to T.~Riemann and S.~Riemersma
for a critical reading of the manuscript.
\newpage


\vspace{3.0cm}

\footnoterule

$^{\dagger}${\footnotesize Deceased}

\twocolumn
\newpage
\begin{center}

\mbox{\epsfig{file=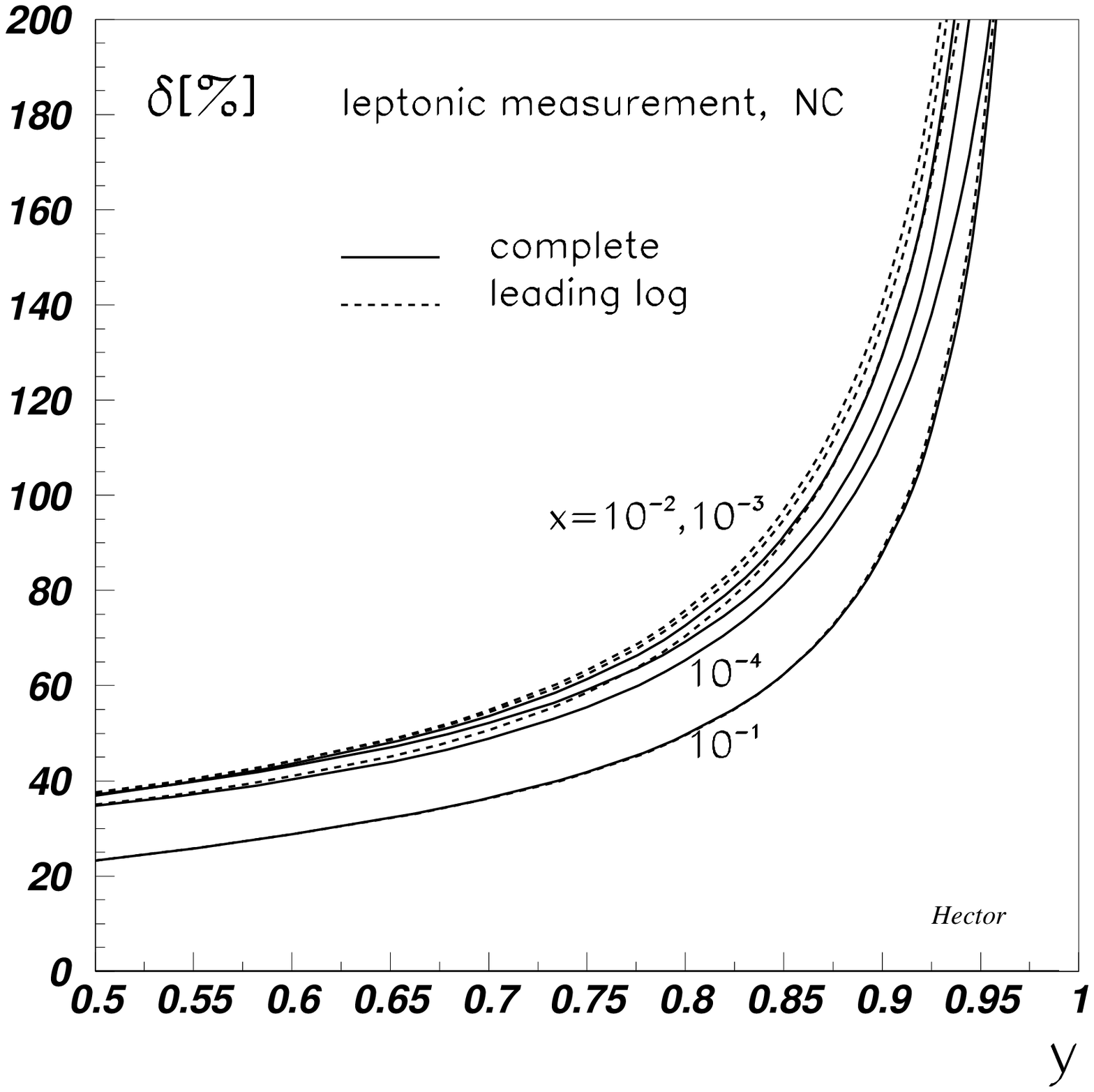,height=9cm,width=8cm}}

\vspace{2mm}
\noindent
\small
\end{center}
{\sf
Fig.~1a:~A comparison of complete and leading log calculations
of RC for NC DIS at HERA for {\it leptonic} variables.
}
\normalsize
\begin{center}
\vspace{5mm}

\mbox{\epsfig{file=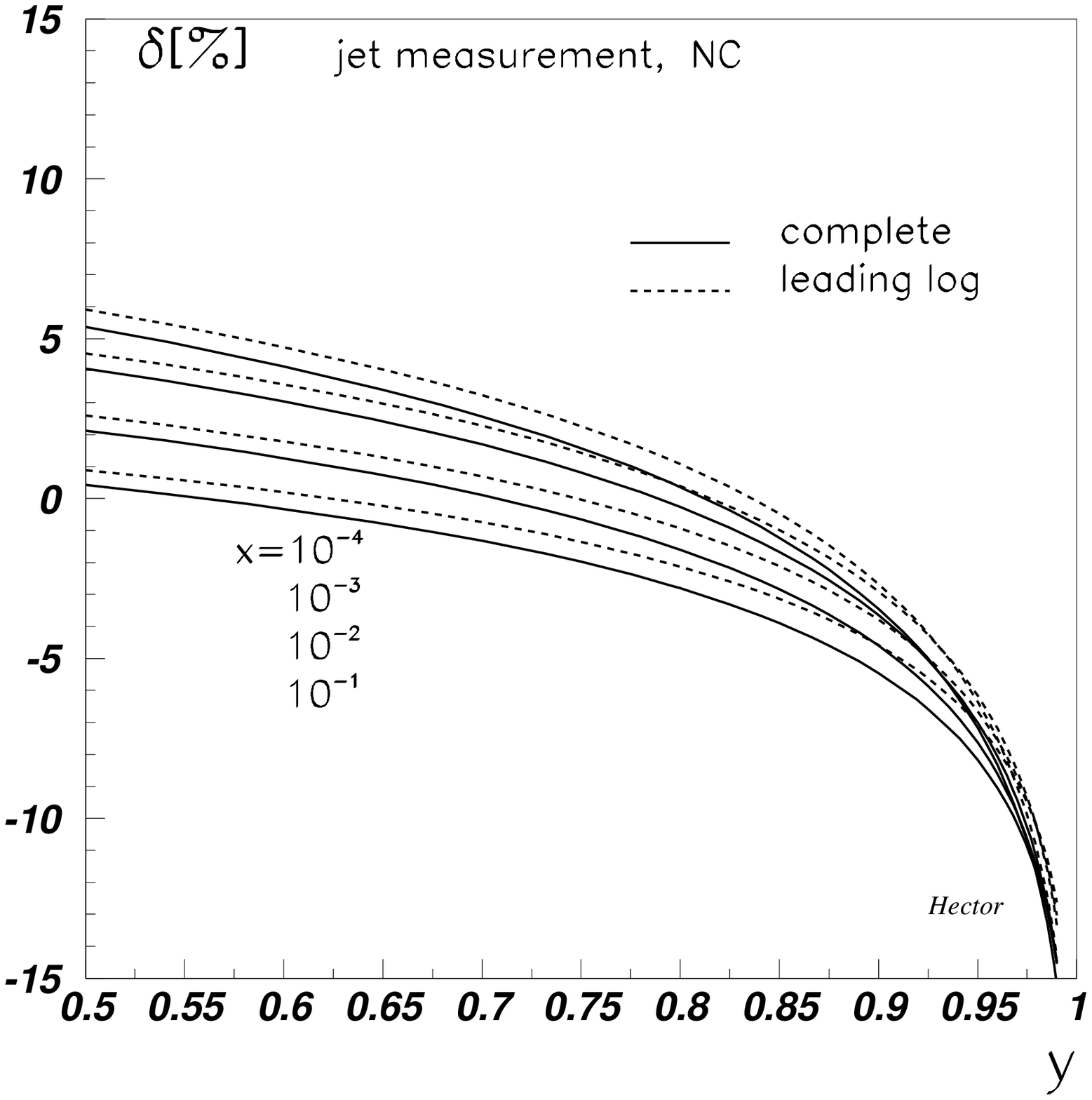,height=9cm,width=8cm}}

\vspace{2mm}
\noindent
\small
\end{center}
{\sf
Fig.~1b:~The same as Fig.1a but for {\it Jaquet-Blon\-del} variables.
}
\normalsize
\newpage
\begin{center}

\mbox{\epsfig{file=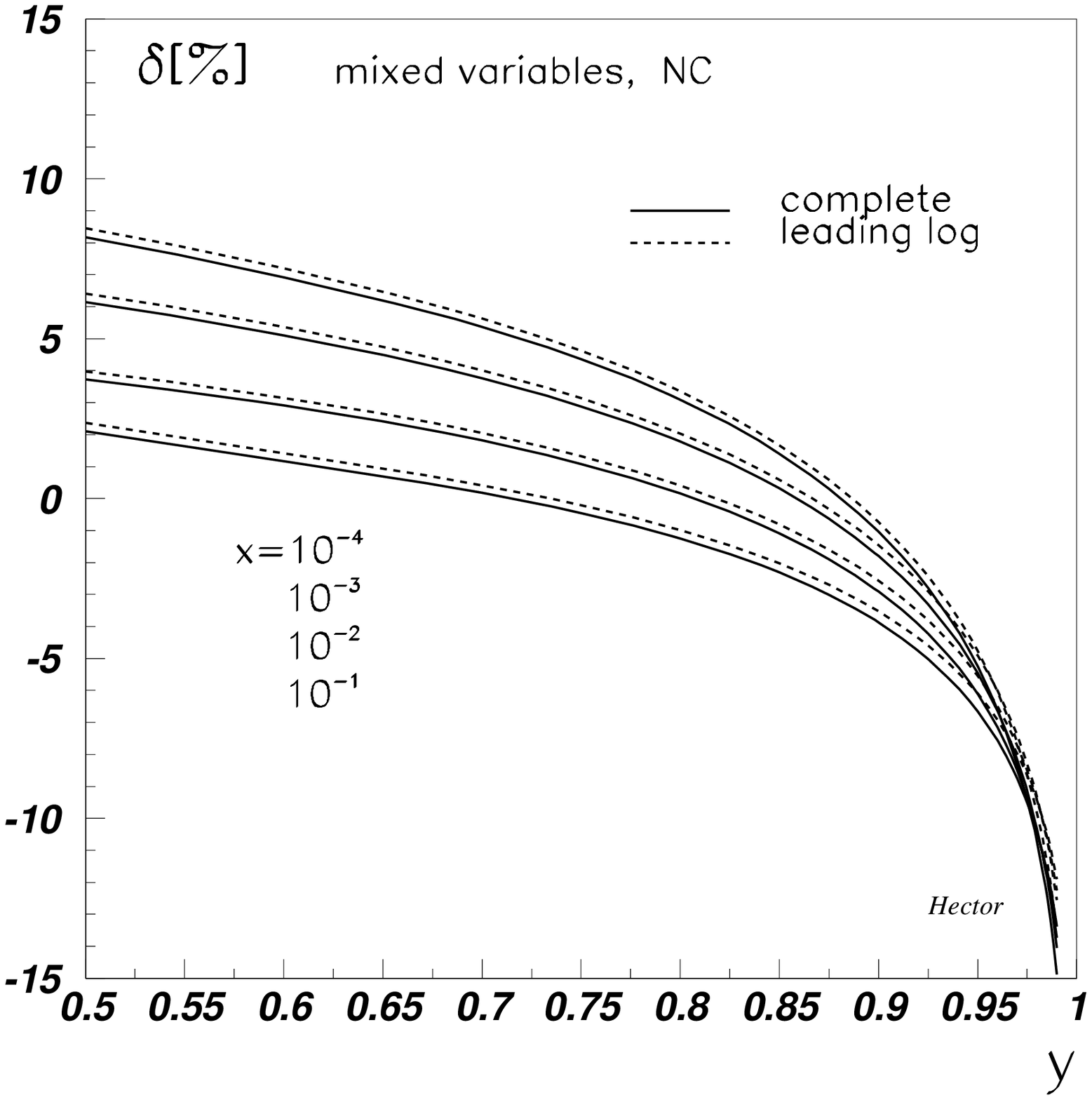,height=9cm,width=8cm}}

\vspace{2mm}
\noindent
\small
\end{center}
{\sf
Fig.~1c:~The same as Fig.1a but for {\it mixed} variables.
}
\normalsize
\begin{center}
\vspace{11mm}

\mbox{\epsfig{file=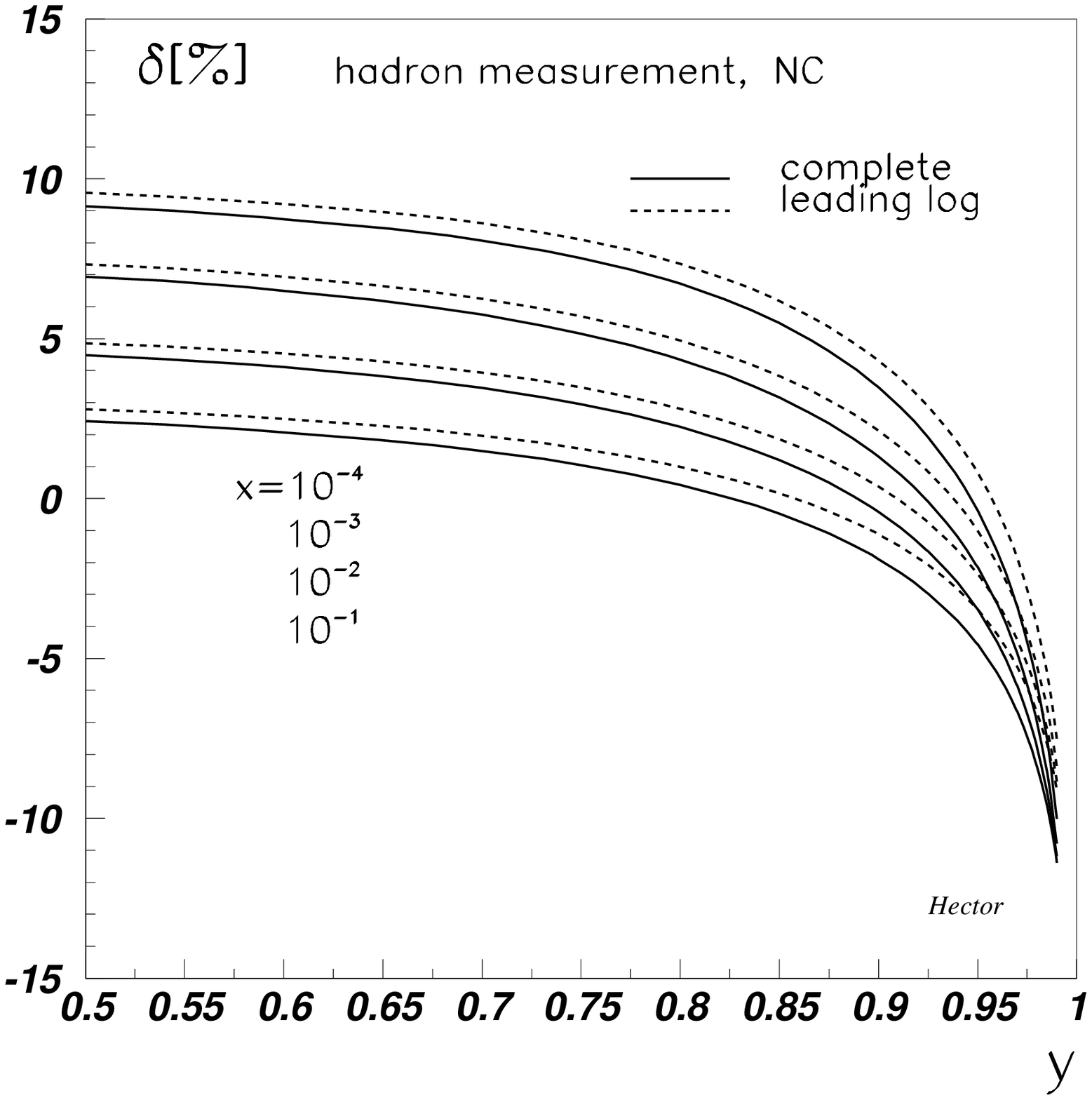,height=9cm,width=8cm}}

\vspace{2mm}
\noindent
\small
\end{center}
{\sf
Fig.~1d:~The same as Fig.1a but for {\it hadronic} variables.
}
\normalsize
\newpage
\begin{center}

\mbox{\epsfig{file=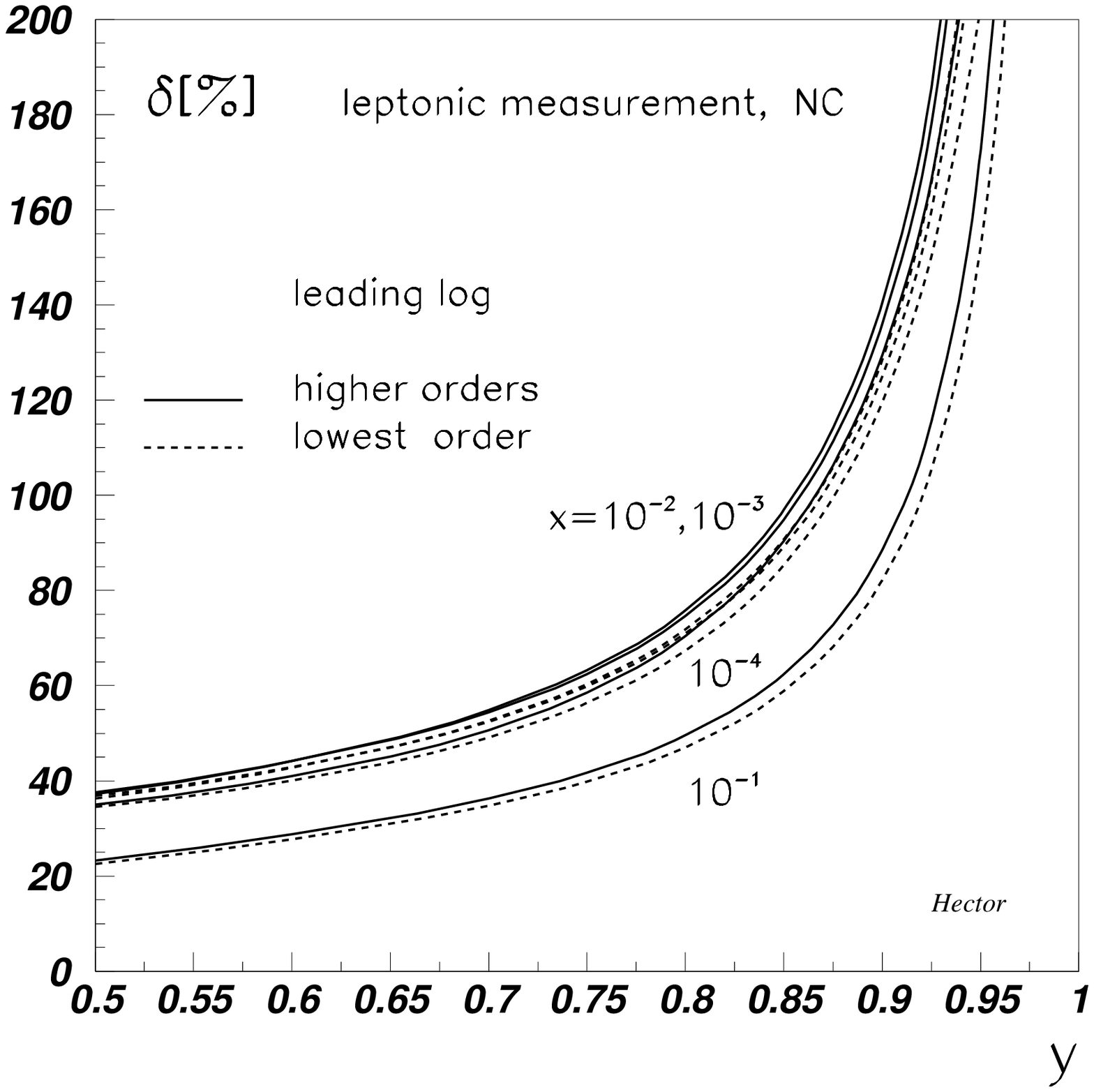,height=9cm,width=8cm}}

\vspace{2mm}
\noindent
\small
\end{center}
{\sf
Fig.~2a:~A comparison
of lowest order with higher order leading log calculations
of RC for NC DIS at HERA for {\it leptonic} variables.
}
\normalsize
\begin{center}
\vspace{5mm}

\mbox{\epsfig{file=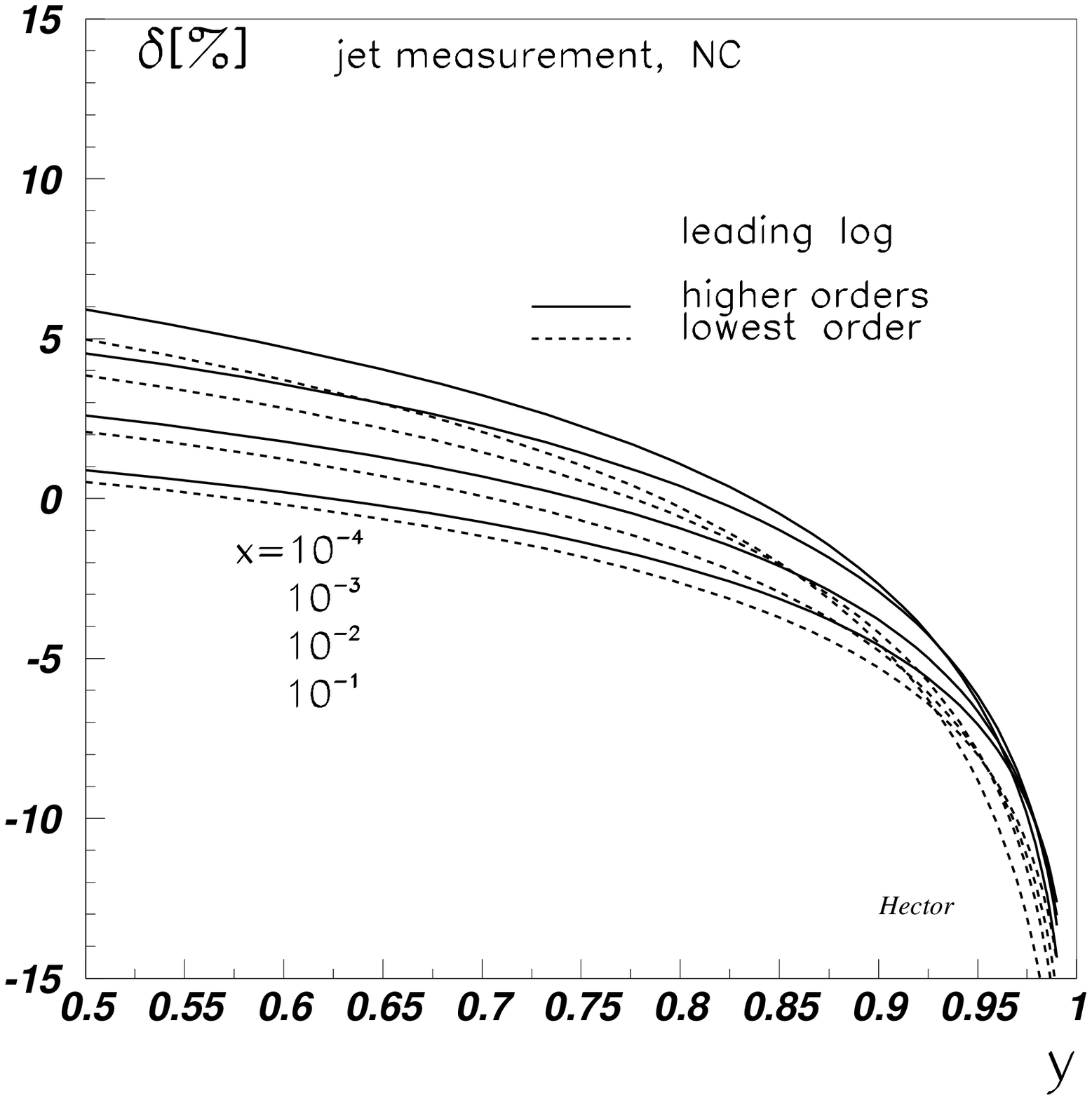,height=9cm,width=8cm}}

\vspace{2mm}
\noindent
\small
\end{center}
{\sf
Fig.~2b:~The same as Fig.2a but for {\it Jaquet-Blon\-del} variables.
}
\normalsize
\newpage
\begin{center}

\mbox{\epsfig{file=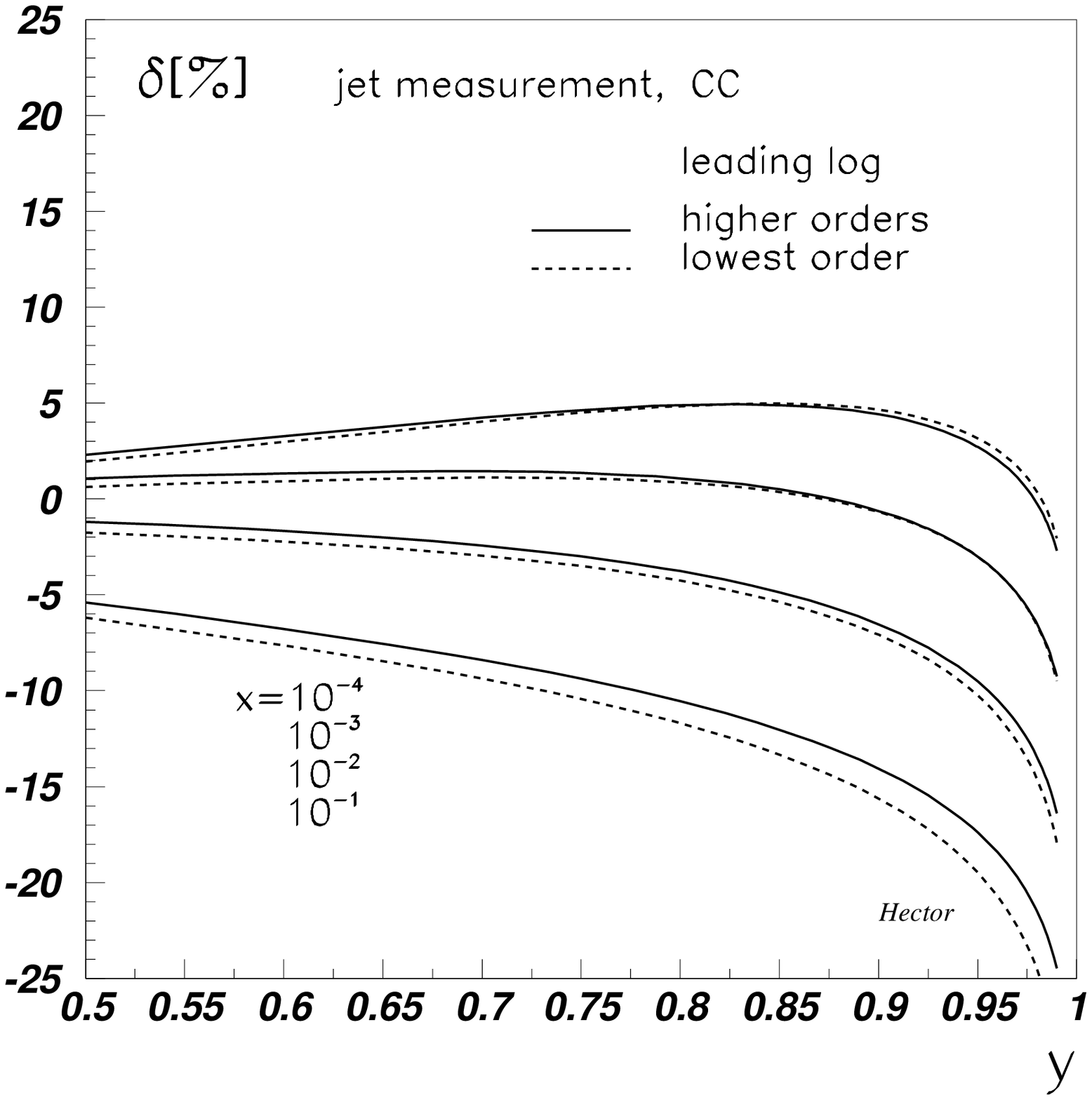,height=9cm,width=8cm}}

\vspace{2mm}
\noindent
\small
\end{center}
{\sf
Fig.~2c:~The same as Fig.2b but for CC DIS.
}
\normalsize
\begin{center}
\vspace{15.25mm}

\mbox{\epsfig{file=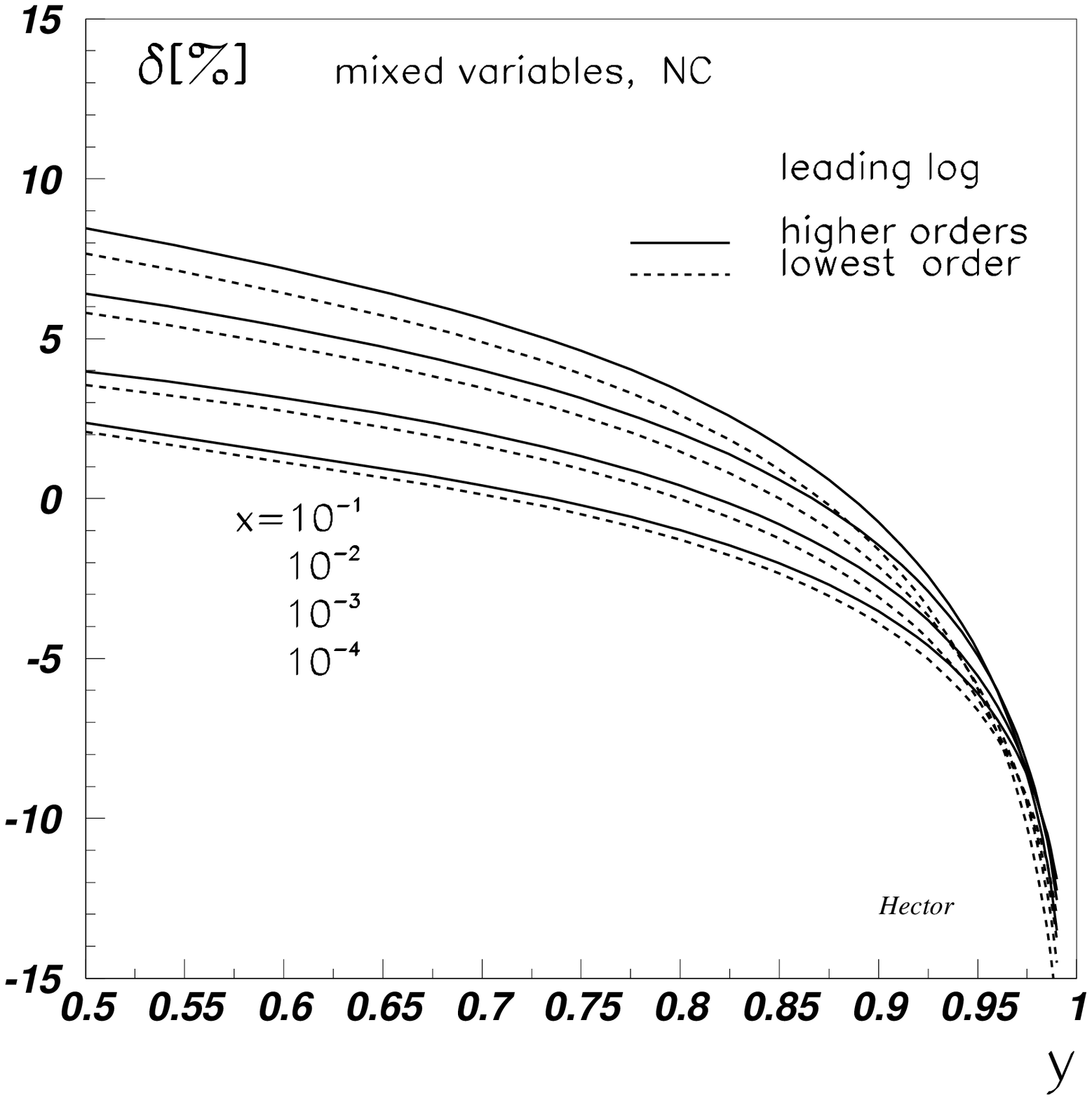,height=9cm,width=8cm}}

\vspace{2mm}
\noindent
\small
\end{center}
{\sf
Fig.~2d:~The same as Fig.2a but for {\it mixed} variables.
}
\normalsize
\newpage
\begin{center}

\mbox{\epsfig{file=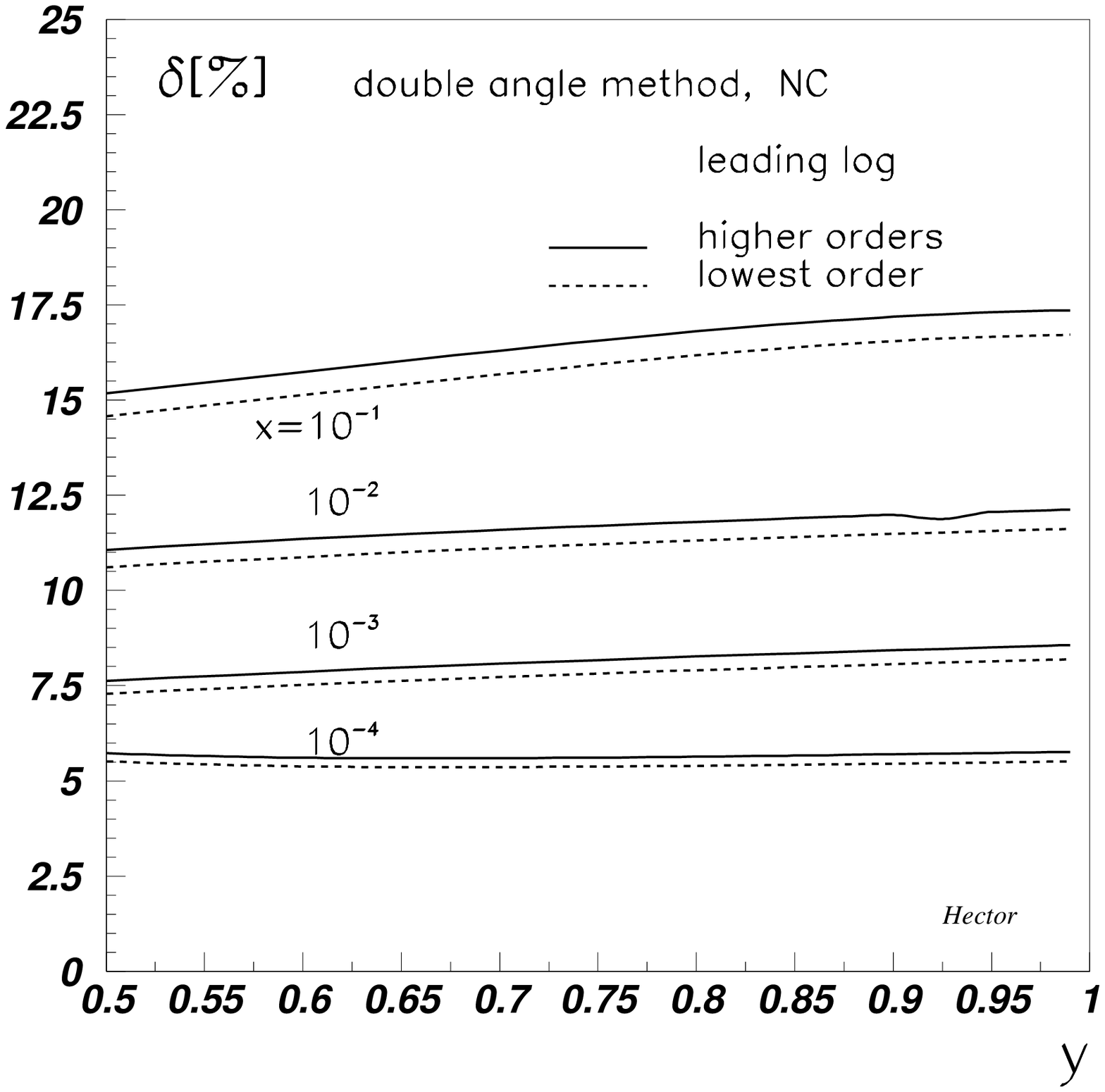,height=9cm,width=8cm}}

\vspace{2mm}
\noindent
\small
\end{center}
{\sf
Fig.~2e:~The same as Fig.2a but for the {\it double angle} method.
}
\normalsize
\begin{center}
\vspace{5mm}

\mbox{\epsfig{file=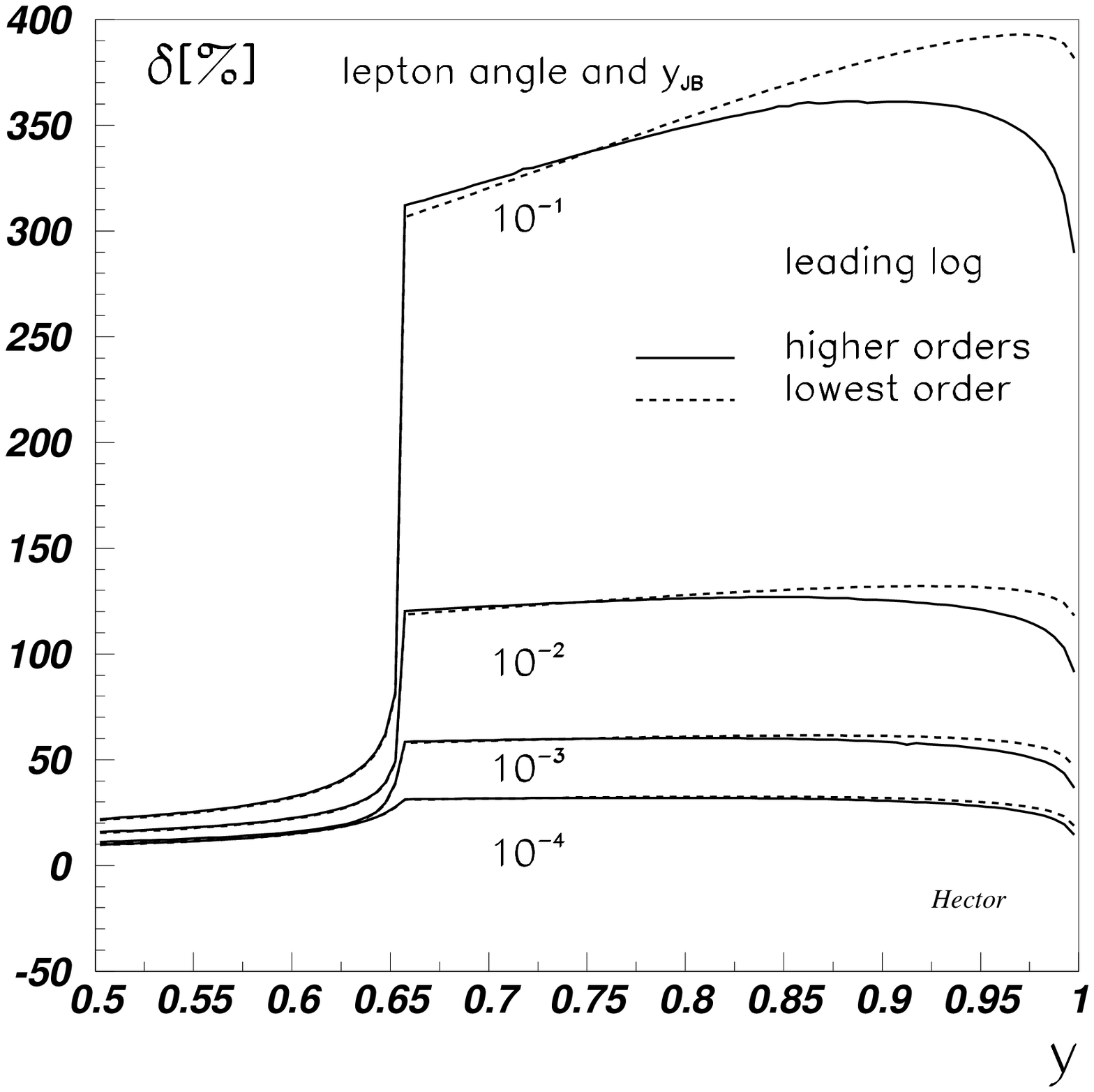,height=9cm,width=8cm}}

\vspace{2mm}
\noindent
\small
\end{center}
{\sf
Fig.~2f:~The same as Fig.2a but for the {\it lepton angle and} $y_{{\rm{JB}}}$
method.
}
\normalsize
\newpage
\begin{center}

\mbox{\epsfig{file=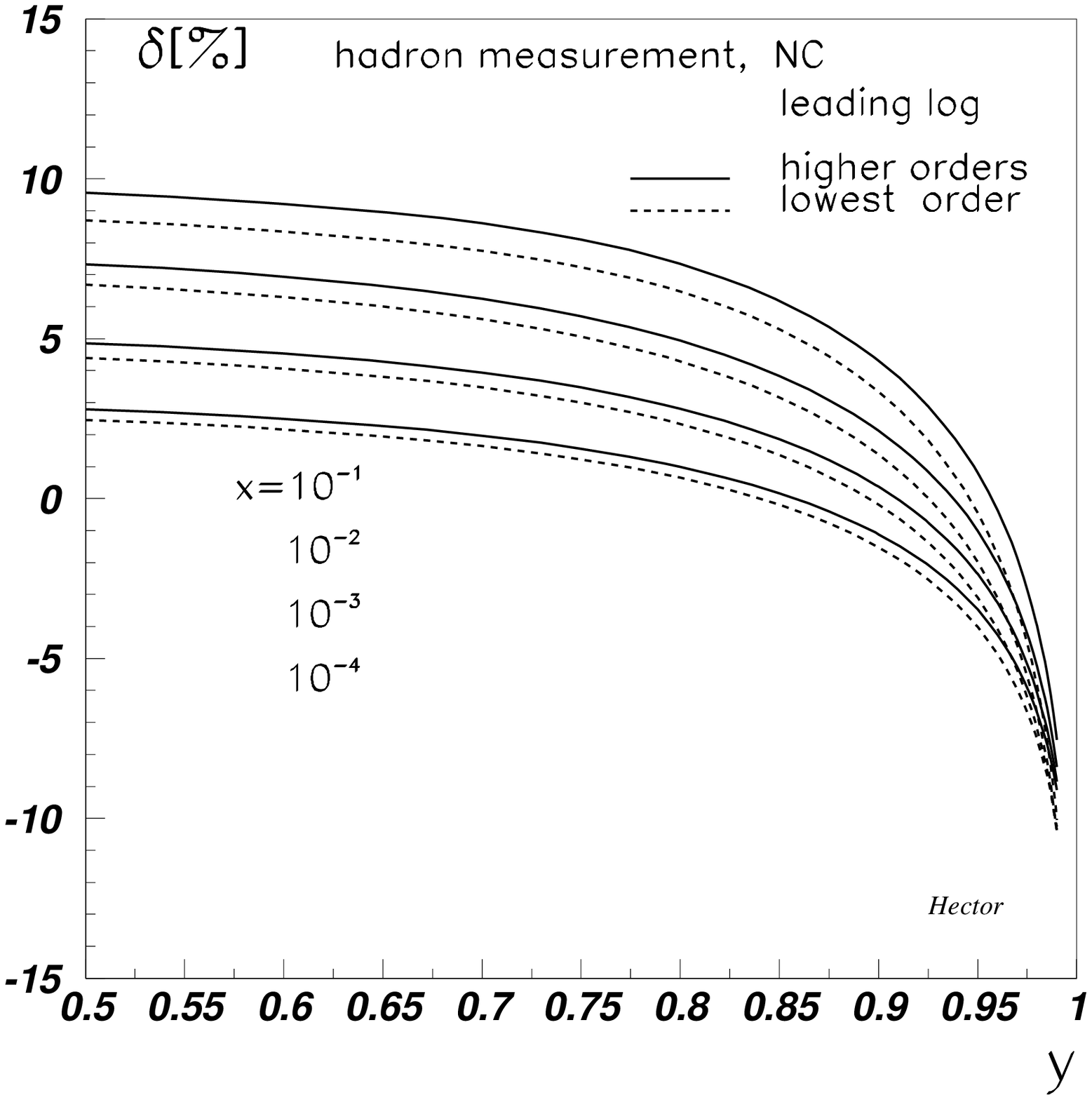,height=9cm,width=8cm}}

\vspace{2mm}
\noindent
\small
\end{center}
{\sf
Fig.~2g:~The same as Fig.2a but for {\it hadronic} variables.
}
\normalsize
\begin{center}
\vspace{5.75mm}

\mbox{\epsfig{file=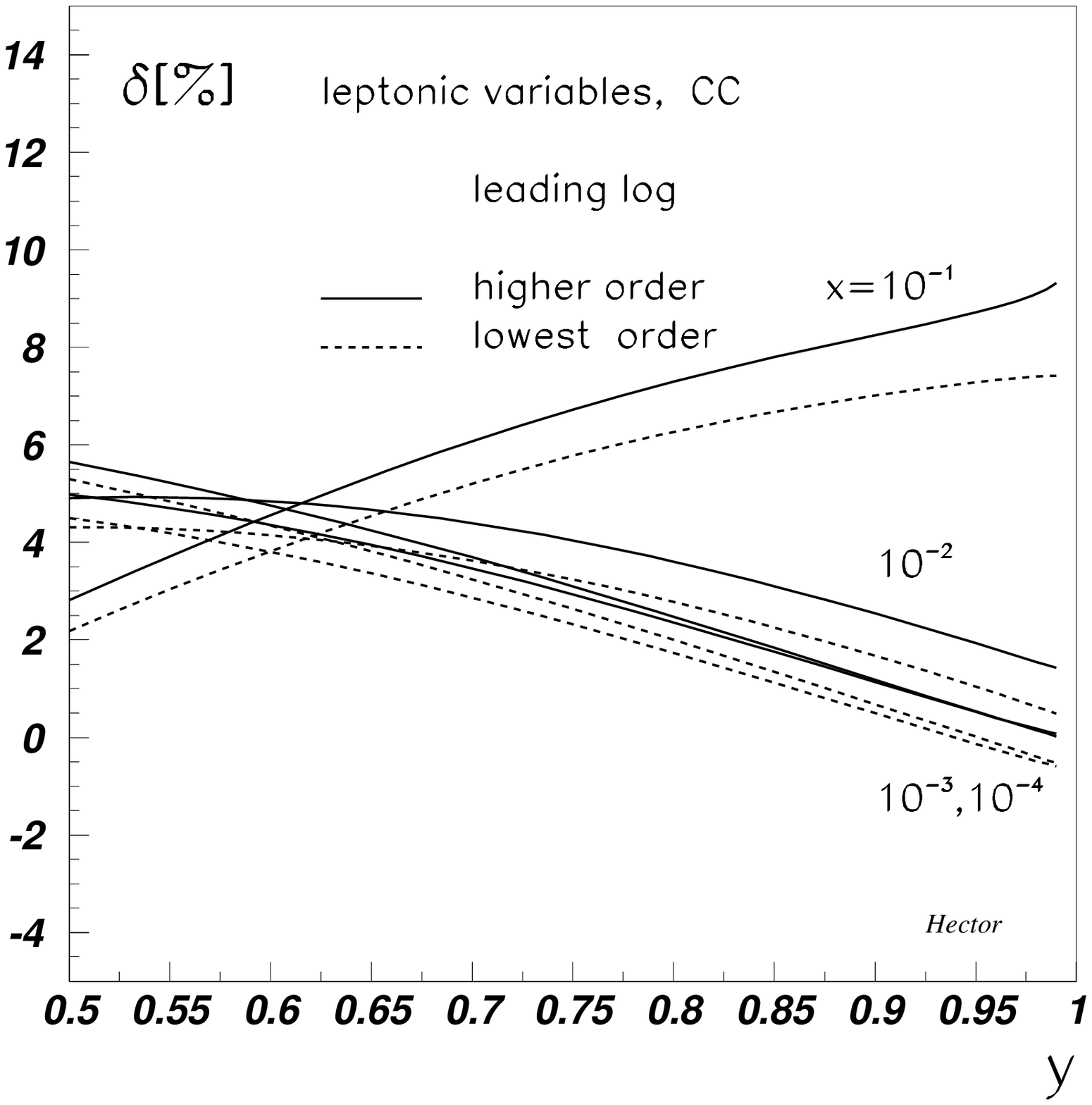,height=9cm,width=8cm}}

\vspace{2mm}
\noindent
\small
\end{center}
{\sf
Fig.~2h:~The same as Fig.2b but for {\it leptonic} variables.
}
\normalsize
\newpage
$\vph$
\vspace{5cm}
\begin{center}

\mbox{\epsfig{file=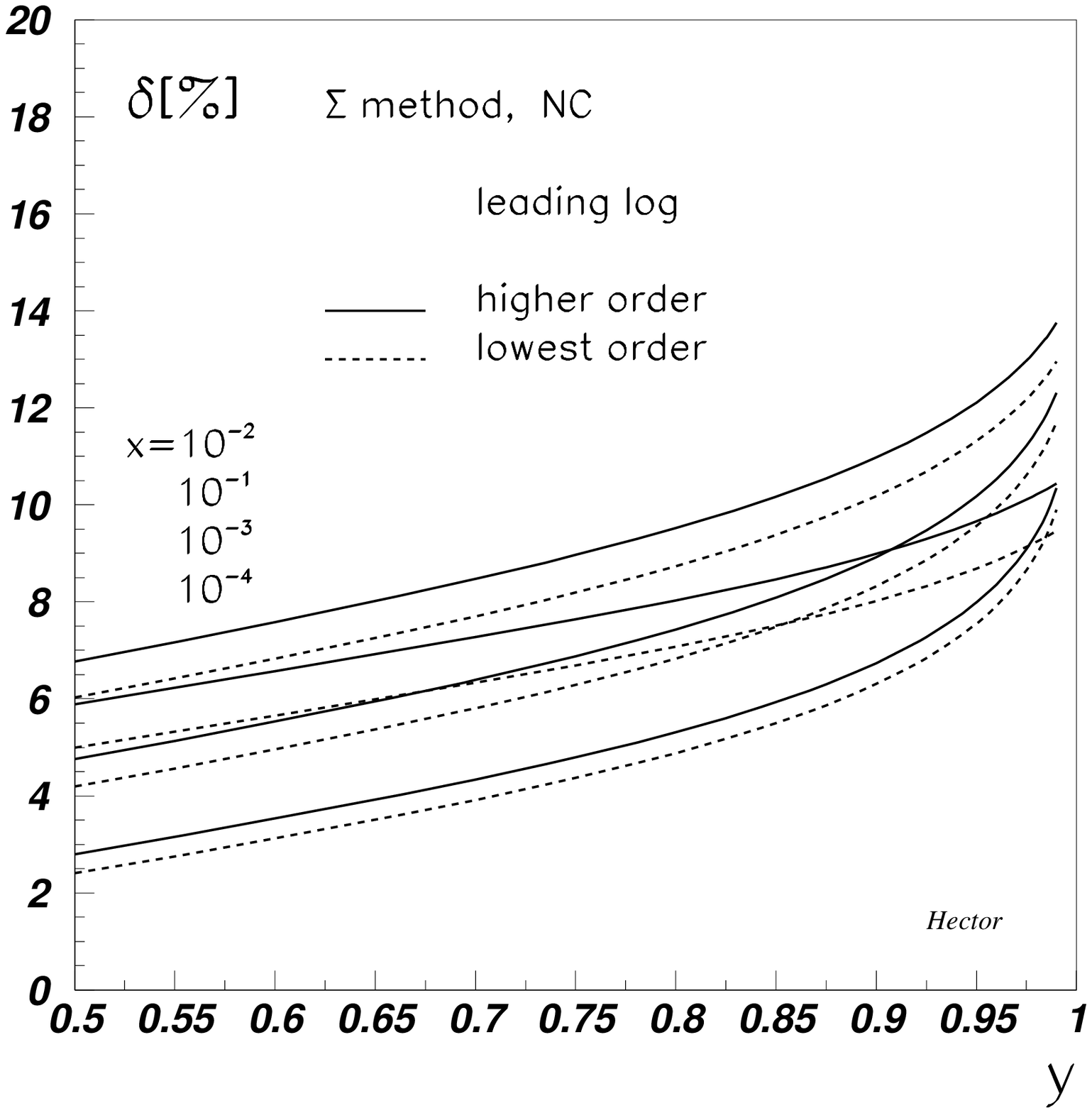,height=9cm,width=8cm}}

\vspace{2mm}
\noindent
\small
\end{center}
{\sf
Fig.~2i:~The same as Fig.2a but for the $\Sigma$ method.
}
\normalsize
\newpage
$\vph$
\vspace{5cm}
\begin{center}

\mbox{\epsfig{file=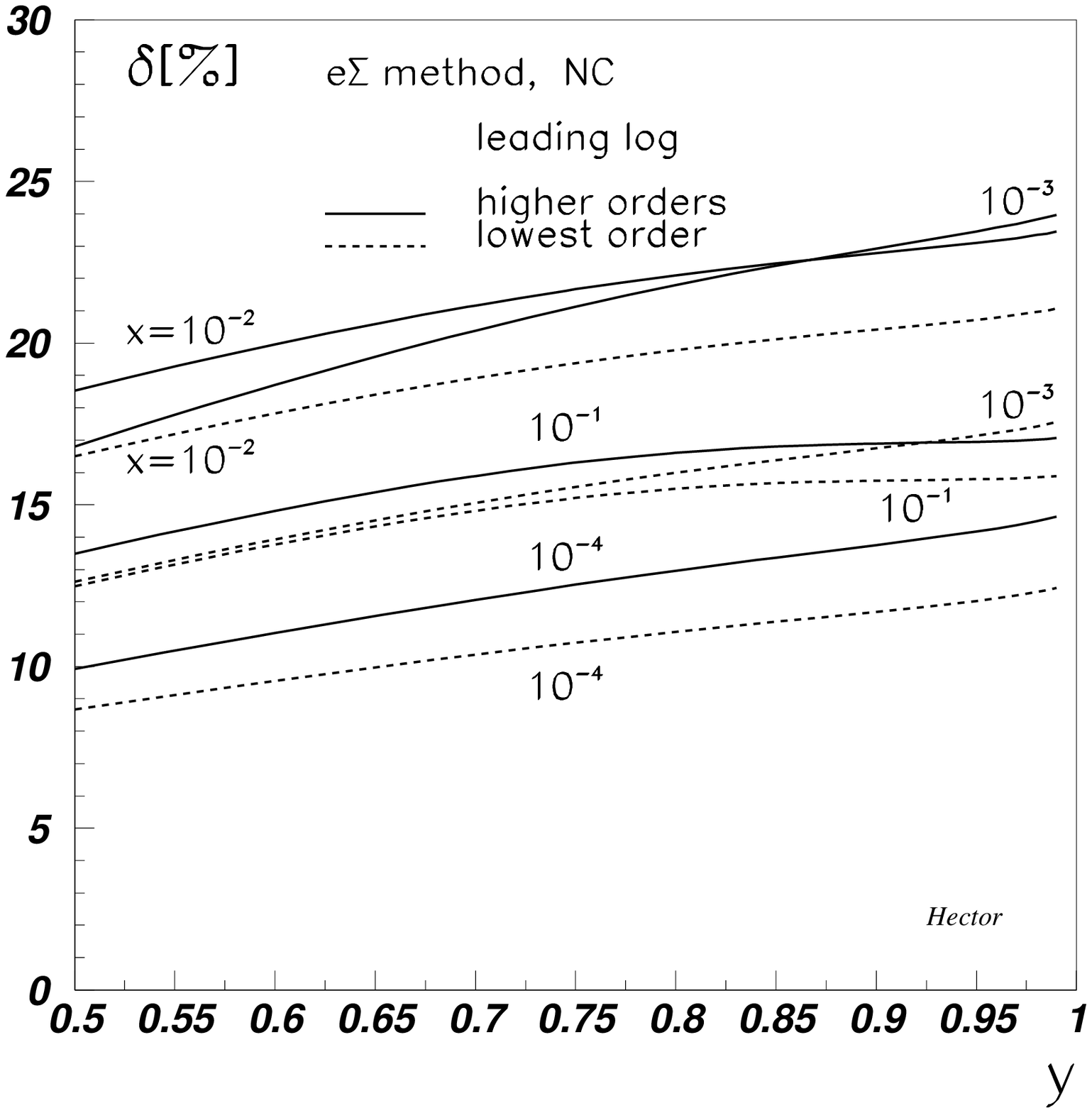,height=9cm,width=8cm}}

\vspace{2mm}
\noindent
\small
\end{center}
{\sf
Fig.~2j:~The same as Fig.2a but for the $e\Sigma$ method.
}
\newpage
\begin{center}

\mbox{\epsfig{file=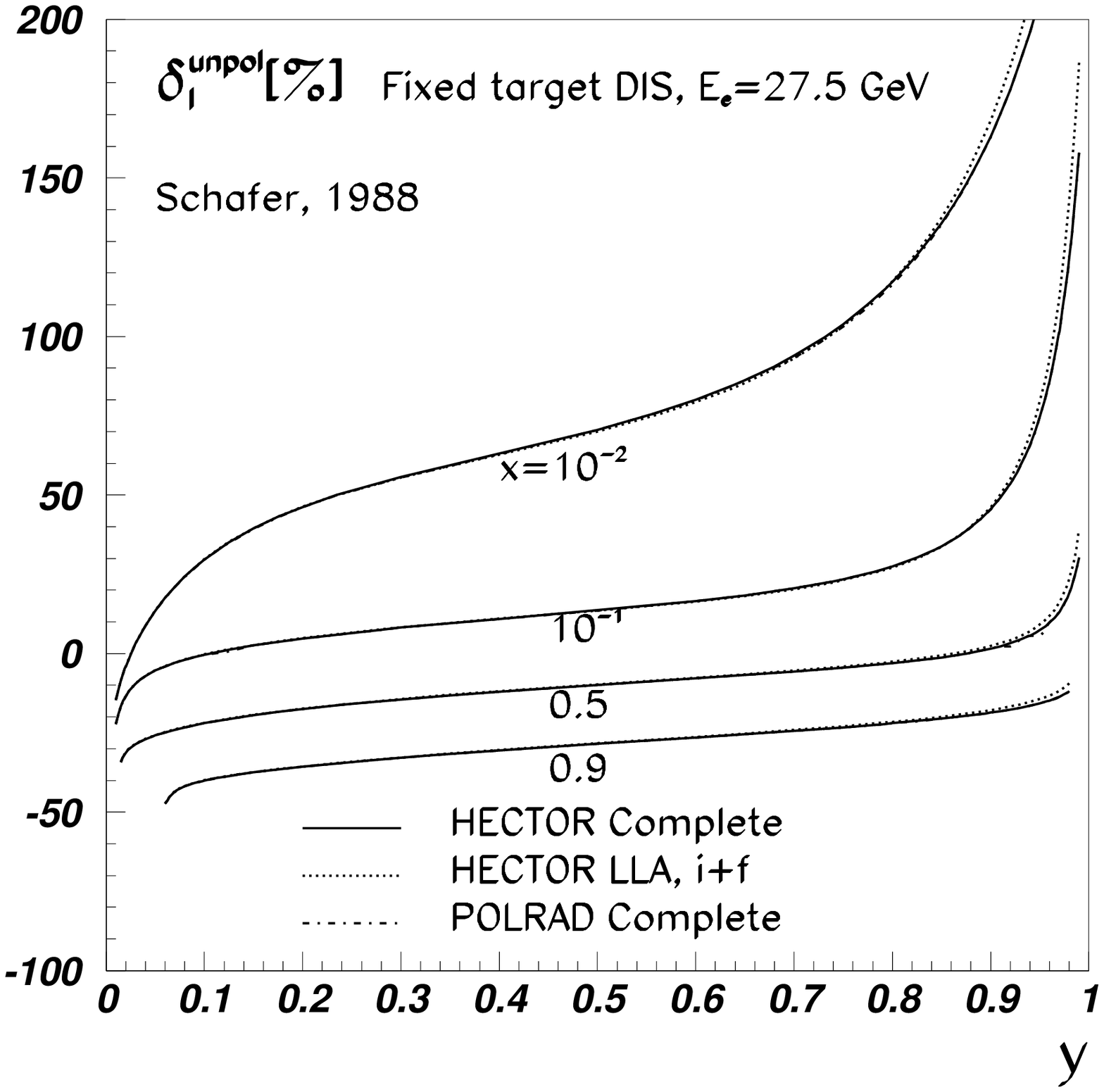,height=9cm,width=8cm}}

\vspace{2mm}
\noindent
\small
\end{center}
{\sf
Fig.~3a:~A comparison of RC between {\tt HECTOR} and {\tt POLRAD}
for NC {\it unpolarized} DIS for {\it leptonic} variables.
}
\normalsize
\begin{center}
\vspace{5mm}

\mbox{\epsfig{file=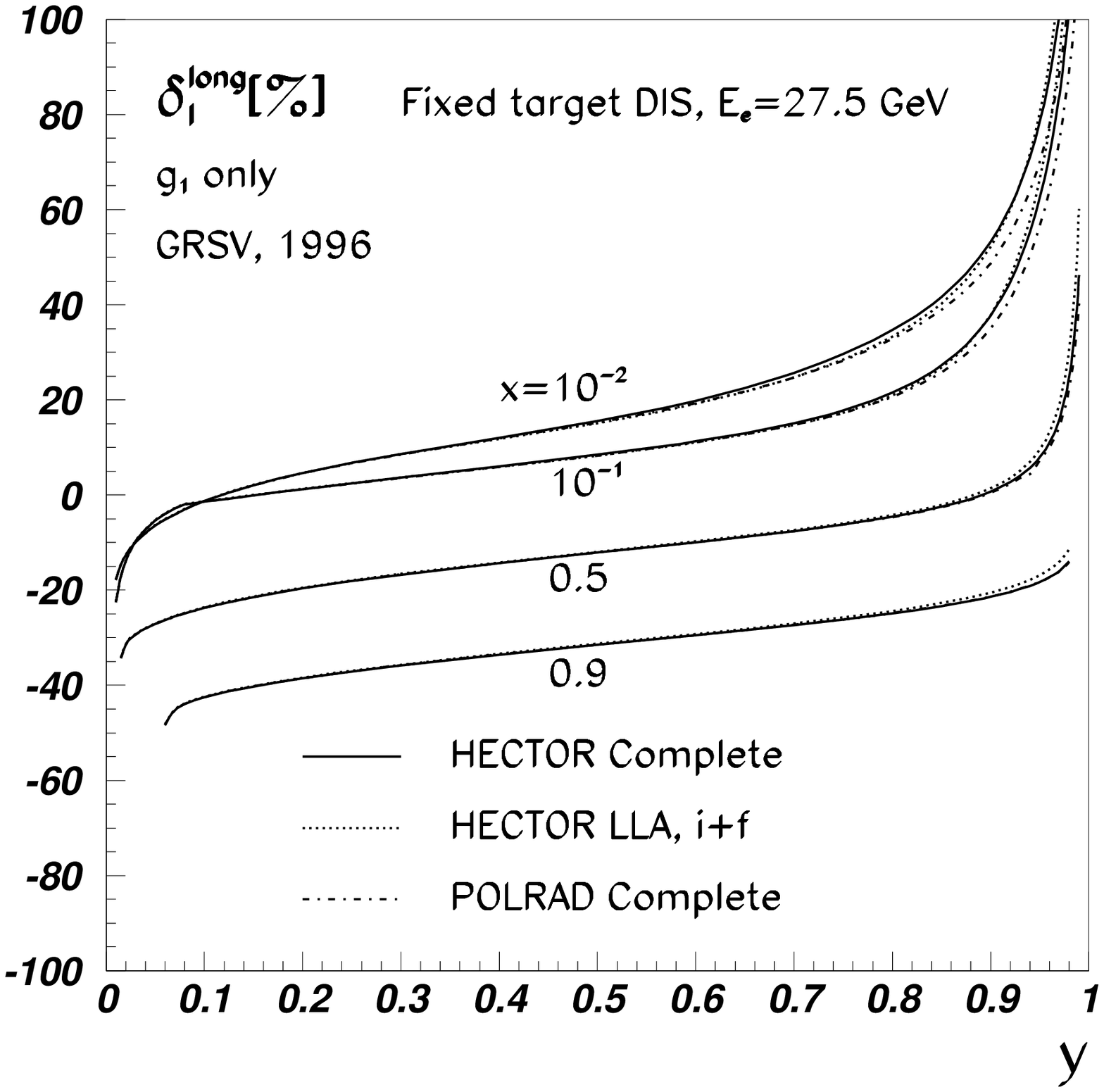,height=9cm,width=8cm}}

\vspace{2mm}
\noindent
\small
\end{center}
{\sf
Fig.~3b:~The same as Fig.3a but for {\it longitudinal} DIS.
}
\normalsize
\newpage
\begin{center}

\mbox{\epsfig{file=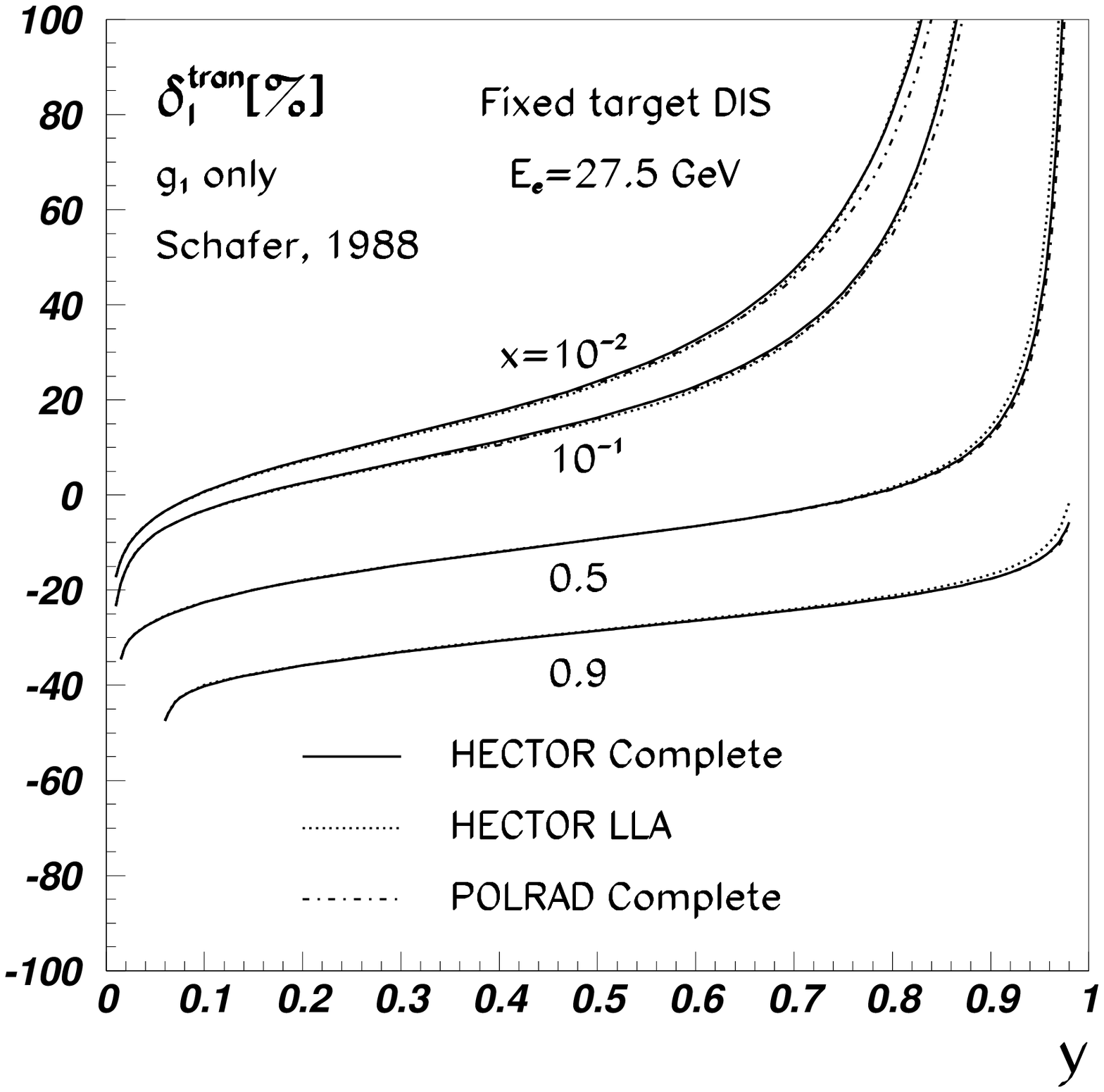,height=9cm,width=8cm}}

\vspace{2mm}
\noindent
\small
\end{center}
{\sf
Fig.~3c:~The same as Fig.3a but for {\it transverse} DIS.
}
\normalsize
\begin{center}
\vspace{10.15mm}

\mbox{\epsfig{file=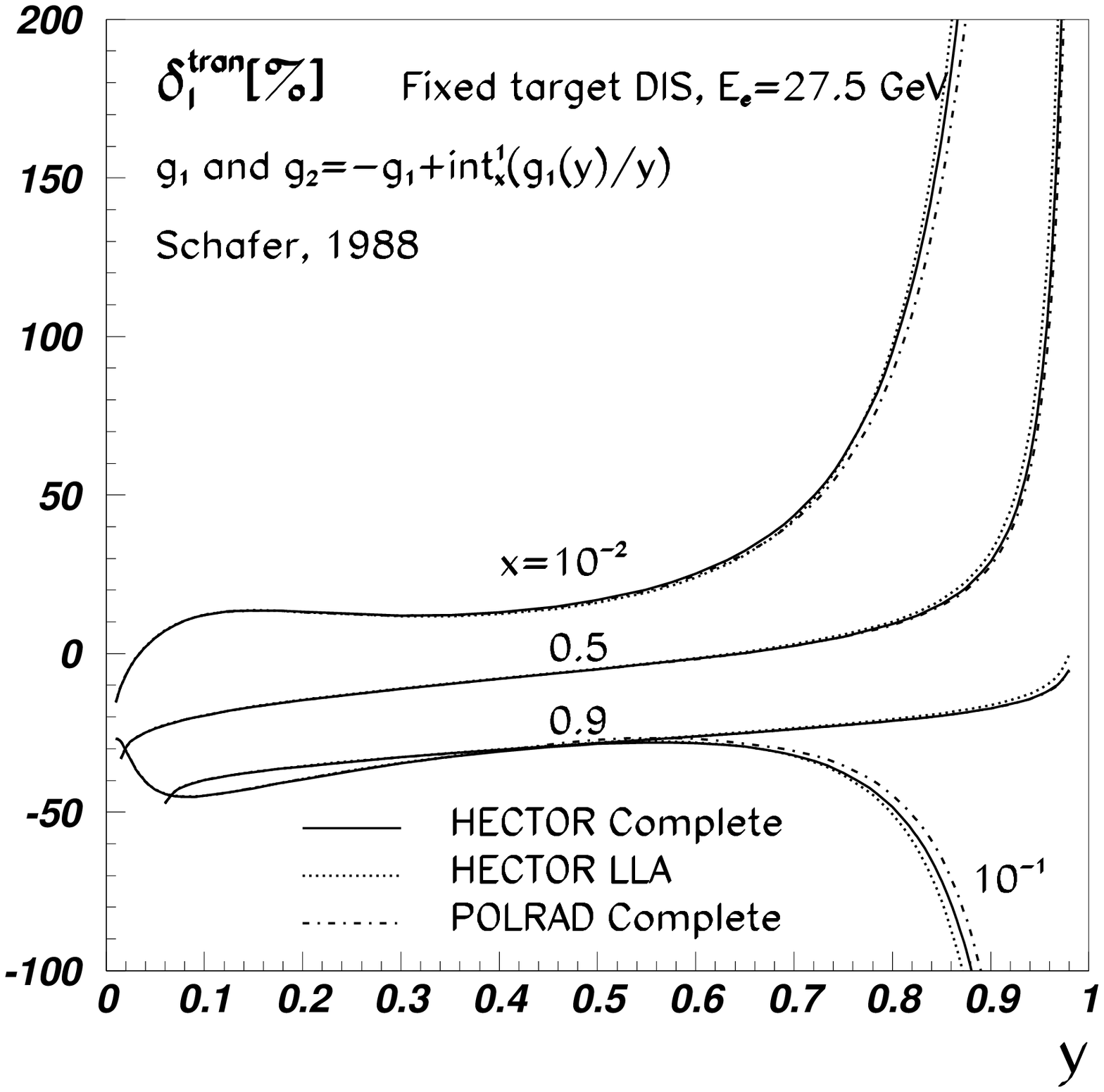,height=9cm,width=8cm}}

\vspace{2mm}
\noindent
\small
\end{center}
{\sf
Fig.~3d:~The same as Fig.3c but for another assumption on $g_2$~\cite{ww}.
}
\normalsize
\end{document}